\documentclass[aps,preprint]{revtex4}
\usepackage{graphicx}% Include figure files

\begin{document}

% define terms

%%%%%%%%%%%%%%%%%%%%%%%%%%%%%%%%%%%%%%%%%%%%%%%%%%%%

%\input{macro}

\newcommand{\eqref}[1]{(\ref{#1})}

\newcommand{\be}{\begin{equation}}

\newcommand{\ee}{\end{equation}}

\newcommand{\bea}{\begin{eqnarray}}

\newcommand{\eea}{\end{eqnarray}}

%%%%%%%%%%%%%%%%%%%%%%%%%%%%%%%%%%%%%%%%%%%%%%%%%%%%

%when use two column
%\twocolumn[\hsize\textwidth\columnwidth\hsize\csname@twocolumnfalse\endcsname\draft

\title{ Modulation of the Curie Temperature in Ferromagnetic/Ferroelectric Hybrid Double Quantum Wells}

\author{N. Kim$^{1}$\footnote{Electronic mail: nmikim@dongguk.edu}, H. Kim$^{2}$, J. W. Kim$^{1}$,
S. J. Lee$^{1}$, and T. W. Kang$^{1}$}
\affiliation{$^{1}$Quantum-functional Semiconductor Research Center,
Dongguk University, Seoul 100-715, Korea\\
$^{2}$Department of Physics, Soongsil  University, Seoul 156-743,
Korea}
%\author{Heesang Kim}
%\affiliation{$^{2}$Department of Physics, Soongsil  University,
%Seoul 156-743, Korea}

\date{December 9, 2005}

\begin{abstract}
We propose a  ferromagnetic/ferroelectric hybrid double quantum well
structure, and present an investigation of  the Curie temperature
($T_{c}$) modulation in this quantum structure.
The combined effects of applied electric fields and spontaneous
electric polarization are considered for a system that consists of a
Mn $\delta$-doped well, a barrier, and a p-type ferroelectric well.
We calculate  the change in the envelope functions of carriers at
the lowest energy subband, resulting from applied electric fields
and switching the dipole polarization.
By reversing the depolarizing field, we can achieve two different
ferromagnetic transition temperatures of the ferromagnetic quantum
well in a fixed applied electric field.
The Curie temperature strongly depends on the position of the Mn
$\delta$-doped layer and  the  polarization strength of the
ferroelectric well.
\end{abstract}

%\\
%Key Words: Magnetic Semiconductor, Hybrid Quantum Structure,
%Ferroelectric Semiconductor, Double Quantum Well \\
%PACS number: 75.50.Pp, 75.75.+a}

%%%%%%%%%%%%%%%%%%%

\maketitle

%]

% "]" should be commented with twocol....

The diluted magnetic semiconductor (DMS) has been generally known as
one of the promising  candidates for  spintronic device materials in
virtue of the coexistence of ferromagnetic and semiconducting
properties in it. Ferroelectric material has also attracted
significant interest because of its promising potential in various
technological applications, such as  binary data storage media in
nonvolatile random access memories due to its spontaneous electric
polarization. In both research fields, many experimental and
theoretical studies have been performed.

Because the spintronic devices should ultimately be operated  at
room temperature, much effort has been focused on increasing the
ferromagnetic transition temperature ($T_{c}$) of DMS above  room
temperature. Among many materials, ZnMnO is considered to have
$T_{c}$ above 300 K with 5$\%$ Mn per unit cell and $3 \times
10^{20}$ holes per cm$^{3}$ according to a theoretical
prediction.\cite{dietl}

Recently, observation of the ferroelectric properties was reported
in Li-doped ZnO bulk samples.\cite{joseph,ferro} The reason for the
ferroelectric property is attributed to the following: when the size
of the dopant Li atom (0.6 \AA) is smaller than the host Zn atom
(0.74 \AA)\cite{joseph}, then the Li atoms can occupy off-center
positions, thus locally inducing electric dipoles, thereby leading
to ferroelectric behavior like PbGeTe\cite{islam} ferroelectric
semiconductor (FES).

The mechanism of DMS ferromagnetism is classified upon materials and
growing techniques. The first class of approach is ferromagnetism
due to the Ruderman-Kittel-Kasuya-Yoshida(RKKY)/Zener indirect
exchange interaction by delocalized holes (hole mediated) based on
the mean-field approximation.\cite{dietl2,macdonald} The second
class of approach is also carrier-induced ferromagnetism as a
results of KKR-CPA-LDA (Korringa-Kohn-Rostoker coherent-potential
approximation and local density approximation) calculations of the
electronic structure of doped DMS alloys.\cite{aki,sato} The third
class of approach suggests the hole hopping mediated ferromagnetism
between polarons having strongly localized charge
carriers.\cite{bhatt,das} And the fourth one is ferromagnetism due
to the ferromagnetic clusters or secondary phases.\cite{m,rao}
Therefore, it is necessary to decide on a case-by-case basis which
mechanism is applicable. In our work, we apply the first class of
approach based on  the mean-field theory for carrier-induced
ferromagnetism in a DMS.\cite{dietl1,bhak,dqw}

Using ideas based on the dependence of  $T_{c}$ of DMS  on the
spatial distribution of magnetic ions, and  envelope functions of
carriers at the lowest energy subband in a confining
potential\cite{bhak,bou,brey}, we model a hybrid double quantum well
(HDQW) system shown in Fig. 1(a). The structure of
ZnO/Zn$_{1-x}$Mg$_{x}$O/Zn$_{1-y}$Li$_{y}$O has the upper p-type ZnO
well with an additional Mn $\delta$-doped layer at the middle of the
well (or at the upper edge of the well). The p-type might be
achieved by the doping of group V \cite{group5,cluster} or group
I\cite{group1} elements. The lower well Zn$_{1-y}$Li$_{y}$O is the
p-type ferroelectric well with spontaneous polarization {\bf P}.
The inverse potential profile of a hole is shown in Fig. 1(b), for
 dipole up and dipole down cases respectively. Because screening
lengths at the interfaces between ZnLiO and ZnMgO are assumed to be
different, the potential profile is asymmetric (different values of
V$_{1}$ and V$_{2}$) in region W$_{2}$.
The dimension of the structure are  chosen: W$_{1}=10$ nm,
W$_{2}=10$ nm, B=5 nm, and the capping layer is $20$ nm. The
confinement potential is $V_{0}$=-263 meV with 20$\%$ of Mg per unit
cell in ZnMgO barriers.\cite{ohtomo}
Because our work is applicable in a regime of low carrier density,
occupying only the  lowest energy  subband of a heavy hole, carrier
concentrations of both wells are in the order of
$10^{11}$/cm$^{2}$\cite{bhak} and the additional band bending due to
carriers is small.\cite{sham2}

We previously demonstrated electric field control of ferromagnetism
in Mn $\delta$-doped asymmetric conventional double quantum wells
 (CDQW)\cite{dqw}, but present structure is different from CDQW
because of hybridizing ferroelectricity. In this work, we can obtain
additional effects from the ferroelectric well due to the reversal
of spontaneous depolarizing fields. We can control the number of
holes around the Mn $\delta$-doped layer by reversing depolarizing
fields as well as  applying electric fields.
The main purpose of this work is to show the possibility of using
ferromagnetic/ferroelectric hybrid structures to modulate $T_{c}$ by
reversing polarization. We can obtain two different ferromagnetic
transition temperatures in the same applied electric field in HDQW.

The Hamiltonian for this system is given by
\bea
H&=&-\partial_{z}\frac{1}{{m}^{\prime}(z)}\partial_{z}+V_{c}(z)-F_{g}z+V_{d}(z).
\eea
Throughout this calculation, we adopt atomic units $
\textit{R}=m_{0}e^{4}/2\hbar^{2}$ for the energy unit, and
$a_{B}=\hbar^{2}/m_{0}e^{2}$ for the length unit, where $m_{0}$ is
the free electron mass. Here $m^{\prime}=m^{*}/m_{0}$ and $m^{*}$ is
the hole effective mass and $F_{g}$ is the carrier charge times for
the applied electric fields. $V_{d}$ is the potential due to the
spontaneous depolarizing field. Because we would like to provide a
qualitative estimation without complication, we ignore the Hartree
and  exchange-correlation interactions among carriers, and
self-consistent solving of electrostatic potential with the Poisson
equation even though the Hartree and exchange-correlation
interactions are related to the effective mass\cite{sham2} and
distribution of carrier spins.\cite{sham1}  The thickness of the
barrier is enough so that we can ignore coupling by tunneling
between the wells. We attempt to solve the eigenvalue problem,
$H\psi(z)=E\psi(z)$, and it becomes
\bea
\frac{1}{m^{\prime}}\partial_{z}^{2}\psi(z)-[V_{c}(z)-F_{g}z+V_{d}(z)-E]\psi(z)&=&0\label{eng}.
\eea
The carrier confinement potential $V_{c}(z)$ is $V_{0}$ inside the
wells and 0 outside the wells. The potential due to dipoles is
$V_{d}(z)=F_{d}\times(z-B/2)+(V_{0}\pm V_{1})$ where $F_{d}$ is the
spontaneous depolarizing field and $V_{1} (>0)$ is the electrostatic
potential at the interface due to screening charges\cite{ye} by the
Thomas-Fermi model of screening.  $\pm$ signs correspond to dipole
left and dipole right cases. This potential profile is shown in Fig.
1(b) with $F_{g}=0$. We know that the general solution for Eq.
\eqref{eng} is a linear combination of Airy functions, $Ai(x)$ and
$Bi(x)$. Thus, we write
\bea \psi(z)&=&C_{1} Ai(\xi)+C_{2} Bi(\xi), \eea
where
\bea \xi&=&\pm(m^{\prime}|F_{g}+F_{d}|)^{1/3} [z-\frac{E-V_{0}\mp
V_{1}+F_{d}\frac{B}{2}}{(F_{g}+F_{d})}].\nonumber \eea
with $\pm=Sgn(F_{g}+F_{d})$ for the region $W_{2}$ and \bea
\xi&=&(m^{\prime}F_{g})^{1/3} [z-\frac{V_{c}-E}{F_{g}}].\nonumber
\eea for elsewhere  with $F_{g}>0$ assumed.
 By using  boundary conditions
%$\psi(z_{L})=\psi(z_{R})=0$ at
%$z_{L}=-B/2-W_{1}-D_{1}$ and $z_{R}=B/2+W_{2}+D_{2}$
and the continuity of wave functions and their derivatives at the
boundaries of each region,
 we can calculate   $C_{1}$ and $C_{2}$ and energy eigenvalues
in the system numerically. In principle, we should include the
effects of the exchange interaction between  Mn ions and carriers on
the carrier wave functions\cite{macdonald,boselli}, but we ignore
these effects because our system has only a very thin Mn
$\delta$-doped layer. Therefore, our wave function is limited to a
system with very thin Mn layers (submonolayer).\cite{sham3} We write
$T_{c}$ in the form\cite{dietl1,bhak}
\bea
T_{c}&=&\frac{S(S+1)J_{pd}^{2}}{12k_{B}}\frac{m^{\ast}}{\pi\hbar^{2}}\int
dz|\psi(z)|^{4}c(z). \label{tc}\eea
Here, $c(z)$ is a magnetic ion distribution function,   $J_{pd}$
 is the exchange integral of carrier-spin exchange interaction,
 and $S$ is a Mn ion spin.  We
calculate  the change in the fourth power of  growth direction
envelope functions, $|\psi(z)|^{4}$, of carriers at the lowest
energy subband in the HDQW as a function of the  applied electric
fields.

In numerical calculations, we choose  physical parameters for
 p-type ZnO/Zn$_{1-x}$Mg$_{x}$O/Zn$_{1-y}$Li$_{y}$O, with
$x=0.2$ and $y=0.05$  The confinement potential $V_{0}=-263$ meV,
and $m_{h}^{\prime}=0.78$ and $1$ for the well and the barrier,
respectively.\cite{ohtomo,mgo}  In our calculation, the exact values
of $J_{pd}$ and $S$ are not required because we calculate the ratio
of $T_{c}$ in Eq. \eqref{tc}. Usually $J_{pd}$ is one of the
important factors in determining the size of $T_{c}$.

Figure 2(a) shows the dependence of the ratio of ferromagnetic
transition temperatures, $T_{c}/T_{c0}$, on bias voltage applied
across the dipole left HDQW, for Mn center-doped (open triangle),
and Mn edge-doped (closed circle) respectively. Here $T_{c0}$ is
$T_{c}$ at $F_{g}=0$ meV/nm for the Mn center-doped CDQW without Li
doping.\cite{dqw}  For Zn$_{0.95}$Li$_{0.05}$O, there are
approximately1.05 dipoles/nm$^{3}$, and they induce a maximum
polarization of approximately 8.77 $\mu$C/cm$^{2}$. While  there is
no depolarizing field (a full screening case) when barriers are
metal, the maximum depolarizing field can be up to
 1.2 eV/nm\cite{note} when  barriers are insulators (no screening charge).
Our case is in between those cases,  the depolarizing field
 $F_{d}=-$0.01 meV/nm and the screening
electrostatic potential $V_{1}$=0.02 meV are used as input
parameters for dipole down FES. Applied electric fields shift the
envelope functions according to the change of potentials, as shown
in  Fig. 2(b). Then,
 the effective hole concentration increases (decreases) in the
 Mn edge-doped layer (center-doped), and
 $T_{c}$ increases (decreases) as a result. When $F_{g}$ is less than
 2 meV, there are few carriers confined at the lowest subband of the
 Mn doped well because of
 the effect of negative (dipole left) $F_{d}$.

Figure 3 shows the dependence of the ratio of ferromagnetic
transition temperatures $T_{c}/T_{c0}$ on bias voltage applied
across the Mn center-doped HDQW for  dipole down (closed circle),
and dipole up (open triangle) respectively. By reversing the
direction of spontaneous polarization, the change in Curie
temperature occurs below $F_{g}$=2 meV. This effect is caused by
asymmetry of electrostatic potential due to screening charges. The
larger asymmetry is, the more effective the reversal is.
Therefore, it is important to fabricate a sample having asymmetric
potential to obtain this result. Both dipole down and dipole up
cases have the same value of $T_{c}/T_{c0}$ above $F_{g}$=2 meV,
because the envelope functions of carriers depend on the potential
profile of the Mn doped well (left side well), which are not
affected by $F_{d}$ as shown in the inset. When the coercive field
of the ferroelectric well is much smaller than the $F_{g}$=2 meV,
the Curie temperature may change at the coercive field. But we do
not take the reversal of polarization due to $F_{g}$ into account,
because we are interested in the regime of $F_{g}$ lower than the
coercive field. The inset shows the change in the potential profile
due to the reversal of dipole polarization in the HDQW as a function
of $z$ at $F_{g}$=4 meV/nm. The potential profiles in the left-hand
side well are the same regardless of the direction of dipoles. The
energies of the lowest subbands are the  same. These energy
eigenvalues are shown in Fig. 4.

Figure 4 shows  energy eigenvalues as a function of  applied
electric fields for HDQW. Solid lines indicate the  upper energy
 limit of the Mn delta-doped well. The energy degeneracy occurs at $F_{g}=0$
 for CDQW. When we compare the dipole right case and
 dipole left case of Fig. 4, the energy degeneracy at $F_{g}=0$
 shifts to the left (dipole right)
 or to the right (dipole left), because the energy levels corresponding to the Mn
 doped well are the same, and only the energy levels of the
  FES well are shifted up (dipole right)
 or down (dipole left) by
 the depolarizing field.
 Therefore, the lowest subband transition occurs in the
 dipole left FES well case in Fig. 4(b). It causes the abrupt
 increase of $T_{c}/T_{c0}$ in Fig. 3.

Figure 5 shows the effect of the depolarized  field strength on the
dependence of the ratio of ferromagnetic transition temperatures
$T_{c}/T_{c0}$. We display three different depolarized field and
apply  the bias voltage  across the Mn edge-doped and dipole left
HDQW. As we expected, they have different transition points
depending on the $F_{d}$ values. All three lines merge into one
because the potential profiles of the Mn doped well are not affected
by $F_{d}$, as in Fig. 3.

In conclusion, we have proposed the DMS/FES hybrid double quantum
well structure. By  using the effects of the spontaneous
depolarizing field from the FES well,  we can modulate the
ferromagnetic transition temperature  of the DMS well in this
system. We calculate the Curie ferromagnetic transition temperature
in terms of its dependence on the envelope functions of carriers at
the lowest energy subband. Through the reversal of the depolarizing
field, we obtain two different ferromagnetic transition temperatures
in an applied electric field. This result  opens the possibility of
using ferromagnetic/ferroelectric hybrid quantum structures for
future multinary spin devices.

\begin{acknowledgments}
This work was supported  by Seoul City and the Korea Science and
Engineering Foundation (KOSEF), through the Quantum-functional
Semiconductor Research Center at Dongguk University.
\end{acknowledgments}

%\end{document}

\newpage

\begin{figure}

\noindent \caption{(a) Schematic diagram of DMS/FES hybrid double
quantum wells. The upper well has a 0.5 monolayer of  Mn
$\delta$-doped layer at the middle of the well (or at the upper edge
of the well). The lower well is Li-doped ZnO  with spontaneous
electric polarization. (b) The potential profile of  the HDQW
structures (dotted line) for a dipole up (left) and dipole down
(right) cases is shown.  The confinement potential is V$_{0}=-$263
meV with 20$\%$ of Mg per unit cell in ZnMgO barriers. V$_{1}$ and
V$_{2}$ are electrostatic potential  due to screening charges at the
interfaces.}\label{Figure1}

\noindent \caption{(a) Dependence of the ratio of ferromagnetic
transition temperatures T$_{c}$/T$_{c0}$ on bias voltage applied
across the dipole left HDQW for Mn center-doped (open triangle) and
Mn edge-doped (closed circle) respectively. The system size is
W$_{1}=10$ nm, W$_{2}=10$ nm, B=5 nm. Here T$_{c0}$ is T$_{c}$ at
F$_{g}=0$ meV/nm for the Mn center-doped conventional DQW without Li
doping. (b) Change in the fourth power of the growth direction
envelope functions of carriers at the lowest energy subband (upper
panel), and the potential profile (lower panel) in the HDQW as a
function of $z$ at F$_{g}$=3 meV/nm (solid) and 7 meV/nm (dashed).}
\label{Figure2}

\noindent \caption{Dependence of the ratio of ferromagnetic
transition temperatures T$_{c}$/T$_{c0}$ on bias voltage applied
across the Mn center-doped HDQW for  dipole left (closed circle) and
dipole right (open triangle) respectively. Inset shows the change of
the potential profile due to the reversal of dipole polarization in
the HDQW as a function of $z$ at F$_{g}$=4 meV/nm.} \label{Figure3}

 \noindent \caption{Energy eigenvalues as a function of  applied  electric
fields for hybrid double quantum wells for (a) dipole right and
 (b) dipole left cases. Solid lines indicates the energy upper
 limit of Mn delta-doped well.} \label{Figure4}

 \noindent \caption{Dependence of the ratio of ferromagnetic transition
temperatures T$_{c}$/T$_{c0}$ on bias voltage applied across the Mn
edge-doped and dipole left HDQW for different depolarized electric
fields.} \label{Figure5}
\end{figure}


\begin{thebibliography}{99}
\bibitem{note} Ref. [2]says that the spontaneous polarization and the
coercive field (E$_{c}$) are reported as 0.044 $\mu$c/cm$^{2}$ and 2
kV/cm in bulk (Zn$_{0.83}$Li$_{0.17}$)O. Here we use the possible
 values by theoretical calculation.
\end{thebibliography}
\end{document}